\documentclass[aps,prb,twocolumn,showpacs,reprint,citeautoscript,amsmath,amssymb,superscriptaddress]{revtex4-1}

\usepackage{graphicx}
\usepackage{dcolumn}
\usepackage{bm}
\usepackage{verbatim}
\usepackage{epstopdf}
\usepackage{xcolor}

\begin{document}

\title{Radiative capture rates at deep defects from electronic structure calculations}

\author{Cyrus E. Dreyer}
\affiliation{Materials Department, University of California, Santa Barbara, California 93106-5050, USA}
\affiliation{Department of Physics and Astronomy, Stony Brook University, Stony Brook, New York 11794-3800, USA}
\affiliation{Center for Computational Quantum Physics, Flatiron Institute, 162 5$^{th}$ Avenue, New York, NY 10010, USA}
\author{Audrius Alkauskas}
\affiliation{Center for Physical Sciences and Technology (FTMC), LT-10257 Vilnius, Lithuania}
\author{John L. Lyons}
\affiliation{United States Naval Research Lab, Washington, DC 20375, USA}
\author{Chris G. Van de Walle}
\affiliation{Materials Department, University of California, Santa Barbara, California 93106-5050, USA}

\date{\today}

\begin{abstract}
  We present a methodology to calculate radiative carrier capture
  coefficients at deep defects in semiconductors and insulators from
  first principles.  Electronic structure and lattice relaxations are
  accurately described with hybrid density functional theory.
  Calculations of capture coefficients provide an additional
  validation of the accuracy of these functionals in dealing with
  localized defect states.  We also discuss the validity of the
  Condon approximation, showing that even in the event of large
  lattice relaxations the approximation is accurate. We test the
  method on GaAs:$V_\text{Ga}$-$\text{Te}_\text{As}$ and
  GaN:C$_\text{N}$, for which reliable experiments are available, and
  demonstrate very good agreement with measured capture coefficients.
\end{abstract}



\maketitle

\section{Introduction}

Optical techniques such as absorption, photoluminescence (PL), PL
excitation spectroscopy, and time-dependent PL are powerful tools for
studying defects in semiconductors and insulators \cite{Davies}.
However, an identification of the microscopic nature of the defects
that give rise to specific optical signatures often requires
quantum-mechanical calculations that address the atomic and electronic
structure at the microscopic level.  Specifically, predictive
calculations of radiative capture rates can be compared with rates
determined from time-dependent PL measurements
\cite{Glinchuk1977,Reshchikov2014,Reshchikov_SR_2016} to provide a
microscopic identification of the defects that give rise to optical
transitions.

Radiative processes may also be involved in defect-mediated
Shockley-Read-Hall (SRH) recombination\cite{SR1952,Hall1952},
particularly in wide-band-gap materials.  Ascertaining whether
radiative recombination channels can be detrimental to device
performance requires a quantitative understanding of radiative capture
rates at deep defects.

In the past, calculations of carrier capture coefficients were based
on analytical models \cite{Bebb1969,Bebb1971,Lucovsky1965,Ridley}.  Such models
do not account for the complexity of the electronic structure of deep
defects, which involves, e.g., strong lattice relaxations that often
break the local symmetry.

In this paper we demonstrate a first-principles implementation for the
calculation of radiative carrier capture rates at defects in
semiconductors and insulators.  We will use two well-characterized
defects as case studies: a Ga vacancy and Te donor complex in GaAs,
\cite{Glinchuk1977} and a carbon substitutional impurity on a
nitrogen site in GaN,\cite{Ogino1980,Reshchikov2014} to show that
calculations based on hybrid density functionals are in excellent
agreement with experimental capture coefficients. We quantify the
errors resulting from key approximations and perform comparisons with
model calculations. First-principles calculations of
radiative capture at a carbon impurity
in GaN were also reported by Zhang {\it et al.}~\cite{Zhang2017}
In our work we present a detailed derivation of the carrier
capture rate and point out differences with the work of Ref.~\onlinecite{Zhang2017} that
are important for quantitative accuracy, as evidenced by comparison with experiment.

\section{Formalism}

\subsection{Radiative capture in semiconductors and insulators}

\begin{figure}
\includegraphics[width=\columnwidth]{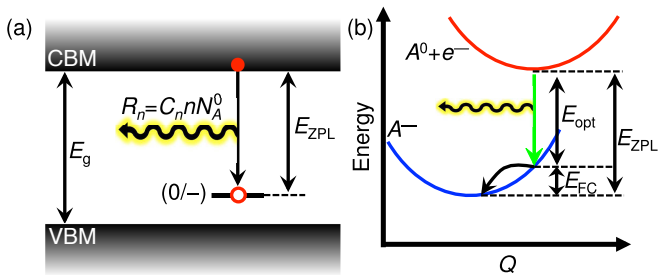}
\caption{\label{sch} Illustration of radiative carrier capture in two
  different representations: (a) band diagram and (b) configuration
  coordinate diagram, for the case of a single acceptor defect $A$.}
\end{figure}

Radiative capture in a material with band gap $E_g$ is illustrated in
Fig.~\ref{sch}(a).  Let us consider a single acceptor defect, $A$,
with a level in the lower part of the band gap. The optical process
consists of the capture of an electron from the conduction band:
$A^{0} + e^{-} \rightarrow A^{-}$. $E_{\text{ZPL}}$ is the zero-phonon
line, given by the position of the $(0/-)$ charge-state transition
level below the conduction-band minimum (CBM). Let $N_A^0$ be the
concentration of acceptors in the neutral charge state, and $n$ the
density of electrons. The rate of the radiative process (i.e., the
number of radiative events per unit time per unit volume) is given by
$R_n=C_nN_A^0n$, where $C_n$ (units: cm$^{3}$s$^{-1}$) is the radiative
electron capture coefficient. A similar equations applies to radiative
capture of holes, with a capture coefficient $C_p$. Determining $C_n$
and $C_p$ is the main goal of the present paper. Instead of
coefficients $C_{\{n,p\}}$, capture cross sections $\sigma$ are often
used. The two are related via $C=v\sigma$, where $v$ is the
characteristic carrier velocity; for non-degenerate carriers $v$ is
the thermal velocity \cite{Ridley}.

The radiative transition can also be represented via a configuration
coordinate diagram [Fig.~\ref{sch}(b)].  The two charge states of the
defect give rise to curves that are displaced along the horizontal
axis because generally they have different atomic configurations (here
projected on a one-dimensional coordinate $Q$).  In the so-called
Franck-Condon approximation, the transition occurs for fixed nuclear
coordinates [see the green arrow in Fig.~\ref{sch}(b)] with energy
$E_\text{opt}$.  After the transition the defect is in a vibrationally
excited state. This state will decay to the equilibrium state on the
picosecond time scale via phonon-phonon interactions, losing the
relaxation energy $E_\text{FC}= E_{\text{ZPL}} - E_{\text{opt}}$,
called the Franck-Condon energy.

The strength of the electron-phonon coupling associated with an
optical transition can be expressed in terms of the Huang-Rhys factor
$S$ \cite{Huang1950}, which quantifies the number of phonons emitted
during the transition. In this work, we will consider defects with
strong electron-phonon coupling ($S\gg 1$). For such defects,
$E_{\text{opt}}$ corresponds to the peak of the PL spectrum
\cite{Alkauskas2012}.  The general formalism to treat optical
transitions in semiconductors is presented in textbooks (
e.g., Ch.~5 of Ref.~\onlinecite{Ridley} or Ch.~10 of
Ref.~\onlinecite{Stoneham}).  Here we will present a derivation of the
capture coefficients specifically adapted to our implementation within
first-principles electronic structure theory, and focusing on the case
of strong electron-phonon coupling.

We will closely follow the reasoning previously applied in deriving
nonradiative capture coefficients \cite{Alkauskas2014} [summarized in
the supplemental material (SM) \cite{SM} section S1], where it was shown
that, for defects in the dilute limit, the capture coefficient can be
expressed as
\begin{equation}
C_n = Vr,
\label{Cn0}
\end{equation}
where $r$ is the capture rate of one electron by one impurity in the
volume $V$; the task is to calculate $r$.

\subsection{Derivation of the capture coefficient}

The wavefunctions describing the defect system are functions of all
electronic $\{x\}$ and ionic $\{Q\}$ degrees of freedom; using the
Born-Oppenheimer approximation, they can be written in the form
$\Psi(\{Q\},\{x\})\chi(\{Q\})$, where $\Psi(\{Q\},\{x\})$ is the
electronic wavefunction (which depends parametrically on $\{Q\}$), and
$\chi(\{Q\})$ is the ionic wavefunction. Let the electronic
wavefunction of the initial (excited) state, which in the case of the
acceptor in Fig.~\ref{sch} is the neutral defect plus the electron in
the conduction band, be $\Psi_i(\{Q\},\{x\})$; the associated ionic
wavefunctions are $\chi_{im}(\{Q\})$, where $m$ denotes the
vibrational state. We will consider only transitions at low
temperature, and therefore initially the system is in the ground
vibrational state ($m=0$).  The corresponding quantities for the final
(ground) electronic state (the negatively charged defect,
Fig.~\ref{sch}) are $\Psi_f(\{Q\},\{x\})$ and $\chi_{fn}(\{Q\})$. The
expressions can be easily generalized to finite temperatures
\cite{Stoneham}.

Optical transitions occur because of coupling to the electric field,
described by the momentum matrix element
$\textbf{P}_{if}(\{Q\})=\langle
\Psi_i(\{Q\})\vert\hat{\textbf{P}}\vert\Psi_f(\{Q\})\rangle$; the
momentum operator 
is $\hat{\textbf{P}}=-i\hbar\sum_j \partial/\partial \textbf x_j$,
where the sum runs over all electrons $j$.  We will use the Condon
approximation (CA) \cite{Lax1952}, in which the dependence on $\{Q\}$
is neglected and the momentum matrix element is taken at a fixed
$\{Q\}$ (which we choose to be the equilibrium geometry of the initial
state); the validity of the CA will be discussed below.

An additional approximation is that multi-electron wavefunctions
$\Psi_{\{i,f\}}$ can be replaced with single-particle Kohn-Sham
orbitals $\psi_i$ and $\psi_f$, with corresponding momentum matrix
element, $\textbf{p}_{if}$.  In the case of electron capture, $\psi_i$
is a perturbed conduction-band state, while $\psi_f$ is a defect
state.  At finite temperature, electrons occupy a thermal distribution
of states with different momenta; in principle, one has to average
over this distribution. For non-degenerate semiconductors at room
temperature these states are very close to the CBM, and thus we will
approximate the initial state to be the CBM.

Within the Born-Oppenheimer approximation and the CA, the luminescence
intensity (number of photons per unit time, per unit energy, for a
given photon energy $\hbar\omega$) is given by \cite{Stoneham}
\begin{equation}
\begin{aligned}
I(\hbar\omega)=& \frac{e^2 n_\text{r}\omega\eta_\text{sp}}{3m^2\varepsilon_0\pi c^3 \hbar}\left|p_{if}\right|^2 \\
& \times \sum_n\left| \left \langle \chi_{i0}|\chi_{fn} \right \rangle \right |^2
\delta \left(E_\text{ZPL}-\hbar \omega_{fn}-\hbar\omega\right) \, .
\label{Lum-int}
\end{aligned}
\end{equation}
$n_{\text{r}}$ is the index of refraction, $\hbar\omega_{fn}$ is the
energy of the final vibrational state (with respect to its ground
state), and $\eta_\text{sp}$ is a factor which accounts for the spin
selection rule ($\eta_{sp}$=1 for a transition from a spin-singlet to
a doublet, $\eta_{sp}$=0.5 for a transition from a doublet to a
singlet or from a triplet to a doublet).  The total recombination rate
is the integral of $I(\hbar\omega)$ over energy $\hbar\omega$:
\begin{equation}
r= \frac{e^2 n_{r}\eta_\text{sp}}{3m^2\varepsilon_0\pi c^3 \hbar^2}\left|p_{if}\right|^2\langle\hbar\omega\rangle \, ,
\label{r1}
\end{equation}
where
$\langle\hbar\omega\rangle = \sum_n\left\vert \langle
  \chi_{i0}\vert\chi_{fn}\rangle\right\vert^2 (E_{ZPL} - \hbar
\omega_{fn})$ is the average energy of emitted photons. In the case of
strong electron-phonon coupling, $\langle\hbar\omega\rangle$ coincides
with the energy of the vertical transition $E_\text{opt}$ [green arrow
in Fig.~\ref{sch}(b)]. For defects studied in this work $S>8$, so we
will make this approximation.

Combining Eq.~(\ref{Cn0}) and Eq.~(\ref{r1}) gives the capture
coefficient if the quantities in Eq.~(\ref{r1}) could be calculated in
a large volume $V$ corresponding to the dilute limit of defects; in
practice, calculations are performed in supercells with much smaller
volumes $\widetilde{V}$. The limited supercell size is not an issue for
describing capture at neutral defects. However, in the case of charged
defects the initial electronic state is not properly described. This
issue can be accounted for by scaling Eq.~(\ref{r1}) by the so-called
Sommerfeld factor $f$
\cite{BonchBruevich1959,Paessler1976,Alkauskas2014}.  In this work,
only neutral centers are considered, so $f=1$.

The final expression for the capture coefficient is:
\begin{equation}
\begin{split}
C_n&= f\eta_{\text{sp}}\widetilde{V}\frac{e^2 n_{\text{r}}}{3m^2\varepsilon_0\pi c^3 \hbar^2}\left|p_{if}\right|^2E_{\text{opt}}
\\
&= \left(5.77\times10^{-17}\right)\left(\widetilde{V}fn_r\eta_{\text{sp}}E_{\text{opt}}\frac{\vert p_{if}\vert^2}{2m}\right)\text{cm}^3\text{s}^{-1}
\label{r2}
\end{split}
\end{equation}
where in the second line we have evaluated the material-independent
parameters (assuming $E_{\text{opt}}$ and $\vert p_{if}\vert^2/2m$ are
expressed in eV, and $\widetilde{V}$ in \AA$^3$) resulting in a simple
formula that can be used to evaluate radiative capture coefficients
based on quantities generated by density functional theory
calculations.
Equation~(\ref{r2}) agrees with the expression used in
Ref.~\onlinecite{Zhang2017}, except for the fact that the spin
selection rules are neglected in that work (e.g., for carbon on the N
site in GaN, this results in an extra factor of two in
Ref.~\onlinecite{Zhang2017}).

\section{Results}

\subsection{Computational details}

We have calculated $E_{\rm opt}$ and $\textbf{p}_{if}$ necessary for
Eq.~(\ref{r2}) using density functional theory with the hybrid
functional of Heyd, Scuseria, and Ernzerhof (HSE) \cite{HSE06}.  The
mixing parameter was chosen to reproduce the experimental band gaps:
0.30 for GaAs \footnote{To account for the effect of spin-orbit
  coupling in GaAs, the value of 0.30 for the mixing parameter is
  chosen to overestimate the band gap by 0.1 eV. Then we rigidly shift
  the VBM up in energy by 0.1 eV.} (giving a band gap of 1.52 eV) and
0.31 for GaN (giving a band gap of 3.50 eV). Ga 3$d$ electrons were
treated as core states. Defect calculations were performed on 216-atom
zincblende supercells for GaAs, and 96-atom wurtzite supercells for
GaN.
When optimizing the defect geometry the Brillouin zone was sampled
with a single special $k$-point (1/4,1/4,1/4) \cite{mp1976}.  Since
for non-degenerately doped GaN and GaAs the electrons participating in
capture originate from the CBM, momentum matrix elements were
evaluated at the $\Gamma$ point. Use of the $\Gamma$ point also
correctly captures the symmetries of the system.  We used the Vienna
\textit{ab-initio} Simulation Package (\textsc{VASP})\cite{vasp1} with
the projector augmented wave (PAW) method \cite{paw};
for the transition matrix elements, the methodology of
Ref.~\onlinecite{Gajdos2006} was used, (i.e., momentum matrix elements
are correctly calculated for the case of nonlocal potentials).
Thermodynamic charge-state transition levels of the defects
($E_{\text{ZPL}}$ in Fig.~\ref{sch}) and $E_{\text{opt}}$ were
calculated using the standard methodology described in
Ref.~\onlinecite{VDWRMP}, where we use the scheme of
Refs.~\onlinecite{Freysoldt2009} and \onlinecite{Freysoldt2011} to
correct for interactions between charged defects and their periodic
images. We use experimental indices of refraction (3.4 for GaAs
\cite{Skauli2003} and 2.4 for GaN \cite{Takahiro1997}, consistently chosen
for energies $E_{\text{opt}}$).

\subsection{Capture coefficients of test-case defects}

We test the methodology on two defects for which extensive
experimental information is available: the complex between a Ga
vacancy and a Te donor on a nearest-neighbor As site in GaAs,
GaAs:$V_\text{Ga}$-$\text{Te}_\text{As}$, \cite{Glinchuk1977} and a
carbon substitutional impurity on a nitrogen site in GaN,
GaN:C$_\text{N}$ \cite{Ogino1980,Reshchikov2014} (see Secs.~S2 and S3
of SM \cite{SM} for details of the
experimental identification).  In both cases, we examine the rate of
electron capture for the neutral charge state.

We calculate the energy of the $(0/-)$ thermodynamic charge-state
transition level\cite{VDWRMP}, at which electron capture occurs, to be
$0.37$ eV above the VBM for GaAs:$V_\text{Ga}$-$\text{Te}_\text{As}$
and $1.02$ eV for GaN:C$_\text{N}$. The calculated and experimental
optical transition energies are given in Table \ref{def_tab}. A
detailed description of the electronic structure of GaN:C$_\text{N}$
and GaAs:$V_\text{Ga}$-$\text{Te}_\text{As}$ is provided in Sec.~S4 of
the SM \cite{SM}.

\begin{table}[t]
\caption{
Calculated (Calc.) and experimentally measured (Exp.) optical transition energies $E_{\rm ZPL}$ and $E_{\rm opt}$ and electron capture coefficients $C_n$ for the two defects in our case studies.
}
\begin{ruledtabular}
\label{def_tab}
\begin{tabular}{c|cc|cc|cc}
  &\multicolumn{2}{c|}{$E_{\rm ZPL}$ (eV)}  &  \multicolumn{2}{c|}{$E_{\rm opt}$ (eV)} & \multicolumn{2}{c}{$C_{n}$ ($10^{-13}$ cm$^3$s$^{-1}$)}  \\
   & Calc. & Exp. & Calc. & Exp.& Calc. & Exp.\\ \hline
  GaAs:$V_\text{Ga}$-$\text{Te}_\text{As}$ & 1.23 & 1.38\footnotemark[1] & 1.02 & 1.18\footnotemark[1] &3.5 &6.5\footnotemark[1]  \\
  GaN:C$_\text{N}$ & 2.48 & 2.57\footnotemark[2] &  2.01 & 2.2\footnotemark[2]& 0.7 &$0.6-1.2$\footnotemark[2]\\
\end{tabular}
\end{ruledtabular}
\footnotetext[1]{Ref.~\onlinecite{Glinchuk1977}}
\footnotetext[2]{Refs.~\onlinecite{Reshchikov2014}, \onlinecite{Reshchikov_SR_2016}}
\end{table}

Our calculated capture coefficients using Eq.~(\ref{r2}) are given in
Table ~\ref{def_tab} along with experimental values. The calculated
value for GaN:C$_\text{N}$ of $C_n=0.7\times10^{-13}$
cm$^{3}$s$^{-1}$, is in good agreement with measurements by Reshchikov
\textit{et al.} \cite{Reshchikov2014,Reshchikov_SR_2016} that yield
values $(0.6-1.2)\times10^{-13}$ cm$^{3}$s$^{-1}$ for radiative
capture coefficients pertaining to yellow luminescence in GaN
(different values are for different samples). Therefore, our
calculations indicate that GaN:C$_\text{N}$ is the likely source of
the yellow luminescence in the samples studied by Reshchikov
\cite{Reshchikov2014}.

Our value for $C_n$ is about a factor of four smaller than the one calculated
in Ref.~\onlinecite{Zhang2017}, which is mainly due to the fact that
$\eta_{\text{sp}}$ is not included in that work.  Additionally, we find a slightly
smaller value of $\vert p_{if}\vert^2$ (0.03 versus 0.05 in
Ref.~\onlinecite{Zhang2017}), and a different value for the
refractive index may have been used.
We note that inclusion of $\eta_{sp}$ (i.e., application of the spin selection rules)
is important and necessary to obtain agreement with experiment.

For GaAs:$V_\text{Ga}$-$\text{Te}_\text{As}$, we find that
$C_n=3.5\times10^{-13}$ cm$^{3}$s$^{-1}$, five times larger than
for GaN:C$_\text{N}$.  Our calculated value is in satisfactory
agreement with the value determined experimentally by Glinchuk {\it et
  al.}\cite{Glinchuk1977} ($C_n=6.5\times10^{-13}$ cm$^{3}$s$^{-1}$).

In order to test convergence with supercell size, we determined $C_n$
for the case of GaN:C$_\text{N}$ using a matrix element
$\textbf{p}_{if}$ calculated in supercells with various sizes. A
72-atom supercell results in $C_n = 0.29\times10^{13}$ cm$^3$s$^{-1}$,
while a 300-atom supercell, gives $C_n = 0.90\times10^{13}$
cm$^3$s$^{-1}$. This indicates that the results for the supercells
used in this study are close to converged.

In Sec.~S5 of the SM \cite{SM}, we compare these HSE results to those obtained
using a generalized gradient functional, demonstrating the necessity
of hybrid functionals for such quantitative accuracy.

\section{Discussion}

\subsection{Accuracy of the Condon approximation}

\begin{figure}
\includegraphics[width=\columnwidth]{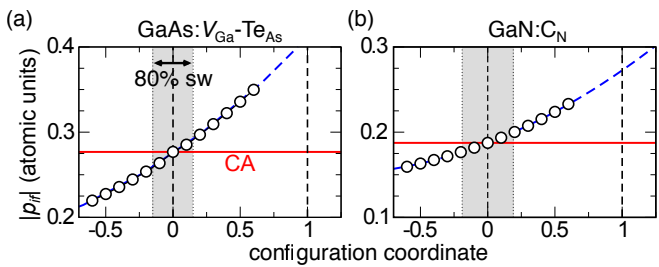}
\caption{\label{FC}
Dependence of the momentum matrix elements $p_{if}$ on configuration coordinate $Q$ for test-case defects (a) GaAs:$V_\text{Ga}$-$\text{Te}_\text{As}$ and (b) GaN:C$_{\text{N}}$.
The blue dashed line represents a quadratic fit to the data.
The gray vertical bar indicates the spread of $Q$ values that gives rise to 80 \% of the spectral weight (sw).
The horizontal line indicates the Condon approximation (CA), in which
the matrix element is taken to be constant.}
\end{figure}

As mentioned above, the derivation of Eq.~(\ref{r2}) relies on the
validity of the CA, which states that the transition matrix element
does not change with configuration coordinate. We now test this
assumption for the two case studies by calculating $\vert p_{if}\vert$
for different $Q$ values (Fig.~\ref{FC}). The generalized
configuration coordinate chosen is a linear interpolation of all
atomic positions between the ground-state structures of the neutral
and negatively charged defects. This choice of $Q$ has been
demonstrated to yield accurate PL lineshapes \cite{Alkauskas2012},
indicating that it is a good approximation for the sum over all
vibrational degrees of freedom.

In Fig.~\ref{FC} the CA is indicated by the red horizontal line,
reflecting the $\vert p_{if}\vert$ values at $Q$=0 (the equilibrium
geometry of the initial state).  In order to estimate the error we
make by using the CA, we must have some measure of the importance that
$Q$ values other than $Q$=0 carry in a full determination of the
transition rate. Such a measure is obtained by calculating the
ground-state vibrational wavefunction in the initial state
[$(A^0+e^-)$ in Fig.~\ref{sch}(b)], the square of which is roughly
proportional to the spectral weight of the optical transition at a
given $Q$.  We then consider the variation of the matrix elements over
the region containing $80\%$ of the spectral weight of the transition
(gray shaded region in Fig.~\ref{FC}).  We see that most of the
spectral weight of the transition is concentrated near the vertical
transition at $Q$=0; this is generally true for defects with large
Huang-Rhys factors, such as the ones considered here.
The matrix element $\vert p_{if}\vert$ varies by $\sim 14\%$ over the
gray range for both defects (Fig.~\ref{FC}), translating into an error
in $C_n$ of less than 14\%, which is acceptable and well within the
experimental uncertainty.  It remains to be seen if the accuracy of
the CA holds for other deep defects.

\subsection{Implications for Shockley-Read-Hall recombination}

The results in Table~\ref{def_tab} provide important information about
the role of radiative capture in defect-assisted SRH recombination
processes.  In Ref.~\onlinecite{Dreyer2016}, it was shown that for
defect densities of $10^{16}$ cm$^{-3}$, capture coefficients larger
than $10^{-10}$ cm$^3$s$^{-1}$ are necessary to result in SRH
recombination rates that would compete with electron-hole radiative
recombination and significantly impact the performance of
light-emitting diodes.  If we use this magnitude as a threshold, we
see that for the defects in Table~\ref{def_tab} the radiative electron
capture rates are much too slow (by three orders of magnitude) to give
rise to detrimental SRH recombination.  We suggest that this
conclusion may be more general.  Based on the character of the
wavefunctions, we expect the optical transition matrix elements for
our case-study defects to be fairly strong, and hence it seems
unlikely that $\vert p_{if}\vert^2/2m$ values for other defects
(including in other hosts) would be orders of magnitude larger.
Furthermore, Eq.~(\ref{r2}) shows that $C_n$ depends only linearly on
$E_{\text{opt}}$.  Both observations indicate that radiative capture
coefficients are unlikely to be high enough to give rise to strong
defect-assisted SRH recombination.

\subsection{Comparison to model calculations}

We now discuss how our methodology and implementation differ from
previous attempts at theoretical descriptions of optical transitions
for defects in semiconductors. Previous methods relied on models for
the defect wavefunction in order to determine $\textbf{p}_{if}$
\cite{Ridley}.  An often-used model for deep defects is the ``quantum
defect'' (QD) model \cite{,Bebb1969,Bebb1971,Lucovsky1965}, where the defect
potential near the core is treated as a square well, while the
long-range part has the form of a Coulomb potential.
It can be shown (see Sec.~S6 of the SM\cite{SM}) that, for capture of
an electron at a neutral acceptor, the QD model results in a
form of the capture coefficient similar to Eq.~(\ref{r2}), but with
the key difference that $\vert p_{if}\vert^2$ is replaced by the
momentum matrix element between the bulk conduction and valence bands
$\vert p_{cv}\vert^2$. The matrix element is then scaled by an
``effective volume'' describing the spatial extent of the defect
wavefunction.  In addition, the QD model uses the zero-phonon
line energy ($E_{\text{ZPL}}$) of the defect instead of
$E_{\text{opt}}$; i.e., the Frank-Condon relaxation energy, resulting
from the coupling with the lattice, is neglected.

We now compare capture coefficients calculated with the QD model
with our full first-principles results.  The equations and parameters
are included in Sec.~S6 of the SM\cite{SM}. We find that for
GaN:C$_\text{N}$, $C_n^{\text{QD}}=0.4\times10^{-13}$
cm$^{3}$s$^{-1}$, which is smaller than our first-principles value
(Table~\ref{def_tab}), and slightly below the experimental range. For
GaAs:$V_\text{Ga}$-$\text{Te}_\text{As}$,
$C_n^{\text{QD}}=10.2\times10^{-13}$, slightly larger than the
first-principles value and overestimating the experimental value.
While for these case studies the model agrees reasonably well with
first-principles results, it is important to emphasize the limited
\emph{predictive} power of models such as the QD
model. First, they require energy levels taken either from experiment
or from first-principles calculations. Second, since
$\vert p_{if}\vert$ is replaced with $\vert p_{cv}\vert$, specific
information about the defect electronic structure is lost, and
assumptions about the character of the defect wavefunction are
required.  In our case studies, the assumption that, as acceptors, the
wavefunctions have valence-band character turns out to be reasonable,
but this will not universally be the case.

\section{Conclusions}

We have demonstrated a methodology for determining
capture coefficients from first principles. For the two case studies
considered, GaAs:$V_\text{Ga}-\text{Te}_\text{As}$ and
GaN:C$_\text{N}$, the calculations give quantitative agreement with
experimental measurements. We also confirmed the validity of the
Condon approximation, a result that can be generalized to all defects
with large values of the Huang-Rhys factor. The procedure outlined in
this work will provide a tool for the identification and
characterization of defects detected by optical spectroscopy, and aid
in identifying the origins and mechanisms of Shockley-Read-Hall
recombination.

\acknowledgments

We thank D. Wickramaratne for advice on the first-principles
calculations, and M. A. Reshchikov (VCU) for fruitful interactions and
for bringing the studies of defects in GaAs to our attention.  Work by
C.E.D. was supported by the US Department of Energy (DOE), Office of
Science, Basic Energy Sciences (BES) under Award No.\ DE-SC0010689.
Work by A.A. was supported the European Union's Horizon 2020 research
and innovation programme under grant agreement No.~820394 (project
A{\sc steriqs}).  J.L.L. acknowledges support from the US Office of
Naval Research (ONR) through the core funding of the Naval Research
Laboratory.  Computational resources were provided by the National
Energy Research Scientific Computing Center, which is supported by the
DOE Office of Science under Contract No.~DE-AC02-05CH11231. The
Flatiron Institute is a division of the Simons Foundation.

\bibliography{RadRecomb}

\end{document}


\title{Supplemental Material: \\
Radiative capture rates at deep defects from electronic structure calculations
}
\author{Cyrus E. Dreyer}
\affiliation{Materials Department, University of California, Santa Barbara, California 93106-5050, USA}
\affiliation{Department of Physics and Astronomy, Stony Brook University, Stony Brook, New York 11794-3800, USA}
\affiliation{Center for Computational Quantum Physics, Flatiron Institute, 162 5$^{th}$ Avenue, New York, NY 10010, USA}
\author{Audrius Alkauskas}
\affiliation{Center for Physical Sciences and Technology (FTMC), LT-10257 Vilnius, Lithuania}
\author{John L. Lyons}
\affiliation{United States Naval Research Lab, Washington, DC 20375, USA}
\author{Chris G. Van de Walle}
\affiliation{Materials Department, University of California, Santa Barbara, California 93106-5050, USA}

\date{\today}

\maketitle

\section{Radiative capture coefficient}

In this section we will derive the expression for the capture
coefficient [Eq.~(1) of the main text], following the reasoning of
Ref.~\onlinecite{Alkauskas2014}.  For the example of electron capture
at a single acceptor, let $V$ be a large volume of the material that
contains $N$ electrons, their density being $n=N/V$. Let $M_A^0$ be
the total number of electron-capturing defects with a density
$N_A^0=M_A^0/V$. The total density of defects is
$N_A^0 + N_A^- = N_A$.
If $\lambda$ is the screening length of the coulomb potential of a
charged defect (or the extent of the short-range potential for a
neutral defect), then the impurity potential essentially vanishes
a few $\lambda$ away from the defect.
We will assume that $\lambda^3 N_A \ll 1$, which means
that the region where the potential of impurities is not negligible is
small compared to the overall volume. This allows us to assume that the
average electron density is equal to the electron density in the
region of negligible impurity potential. Let us also assume that the
electron density is low enough ($n\ll\lambda^{-3}$) so that one
electron interacts with only one impurity at a time (i.e., the
probability of finding two electrons in the volume $\lambda ^3$ is
negligible).

Let $r$ be the capture rate of one electron by one impurity in the
volume $V$. The capture rate of $N$ electrons at $M_A^0$ defects will
thus be $\gamma_n=rM_A^0N$, which we can rewrite as
\begin{equation}
  \label{gammaV}
  \gamma_n/V = (Vr) (M_A^0/V) (N/V) = (Vr) N_A^0n.
\end{equation}
As we state in the
main text, the rate of the radiative process (i.e., the number of
radiative events per unit time per unit volume) is given by $R_n=C_nN_A^0n$, where
$C_n$ is the radiative electron capture coefficient. By comparing this expression with
Eq.~(\ref{gammaV}) and noting that $\gamma_n/V=R_n$, we get:
\begin{equation}
C_n = Vr,
\label{Cn0}
\end{equation}
where $r$ is the capture rate of one electron by one impurity in the
volume $V$.

\section{Experimental determination of capture coefficients} 

Experimentally, values for the radiative capture coefficient can be
determined from time-dependent PL measurements under well controlled
conditions
\cite{Glinchuk1977,Reshchikov2005,Reshchikov2014,Reshchikov_SR_2016}. Let
us consider an $n$-type semiconductor (with electron concentration
$n$) containing the deep acceptor as depicted in Fig.~1 of the main text. The
ground state of the defect is the $-1$ charge state, since the Fermi
level is above the $(0/-)$ charge-state transition level. Thus, in
order to capture an electron radiatively, the defect has to capture a
hole first. Let us first assume that hole capture, which occurs
nonradiatively, is a much faster process than radiative capture. In
this case photo-generated holes will be quickly trapped at defects,
and the time dependence of the luminescence will be determined by the
rate at which neutral acceptor defects are re-filled with electrons:
$r=C_nn$ (units: s$^{-1}$). The radiative lifetime $\tau=1/r$ is
therefore:
\begin{equation}
\tau = \frac{1}{C_nn}.
\label{tau}
\end{equation}
$\tau$ is independent of the defect concentration
\cite{Reshchikov_SR_2016,Reshchikov2005,Reshchikov2014}. The electron
density $n$ can be measured independently, and thus time-dependent PL
measurements provide a direct measurement of capture coefficients
$C_n$. The error bar in the experimental determination of $C_n$ stems
mainly from inaccuracies determining $n$ \cite{Reshchikov2005}.

In more general situations $C_n$ is extracted from the temperature
dependence of radiative decay rates, as recently described by
Reshchikov {\it et al.} \cite{Reshchikov_SR_2016}. In most cases these
decay rates follow closely the behavior deduced from a rate-equation
model, which yields an accurate determination of $C_n$.

\section{Experimental identification of the defects in the case studies}

\subsection{GaAs:$V_\text{Ga}$-$\text{Te}_\text{As}$ }

GaAs defect luminescence lines occurring around 1.2 eV have long been
associated with a gallium vacancy bound to a donor (either S, Se, and
Te on the As site, or C, Si, Ge, and Sn on the Ga site)
\cite{Williams1968}. The involvement of $V_{\text{Ga}}$ was postulated
based on the observation that the lines were absent in GaAs samples
grown from a gallium solution (gallium-rich conditions)
\cite{Queisser1966}. The lines were also absent in heavily Cd- or
Zn-doped samples, which was attributed to the fact that these
acceptors occupy Ga sites \cite{Williams1968}; in fact, this reduction
was more likely due to the lowering of the Fermi level, which
increases the formation energy of $V_{\text{Ga}}$ and hence lowers the
$V_{\text{Ga}}$ concentration.  In addition, the intensity of the
emission increased with increasing donor concentration, again
supporting the involvement of $V_{\text{Ga}}$ \cite{Williams1968},
since the formation energy of $V_{\text{Ga}}$ decreases as the Fermi
level moves up.

\begin{figure}[h]
\includegraphics[width=8cm]{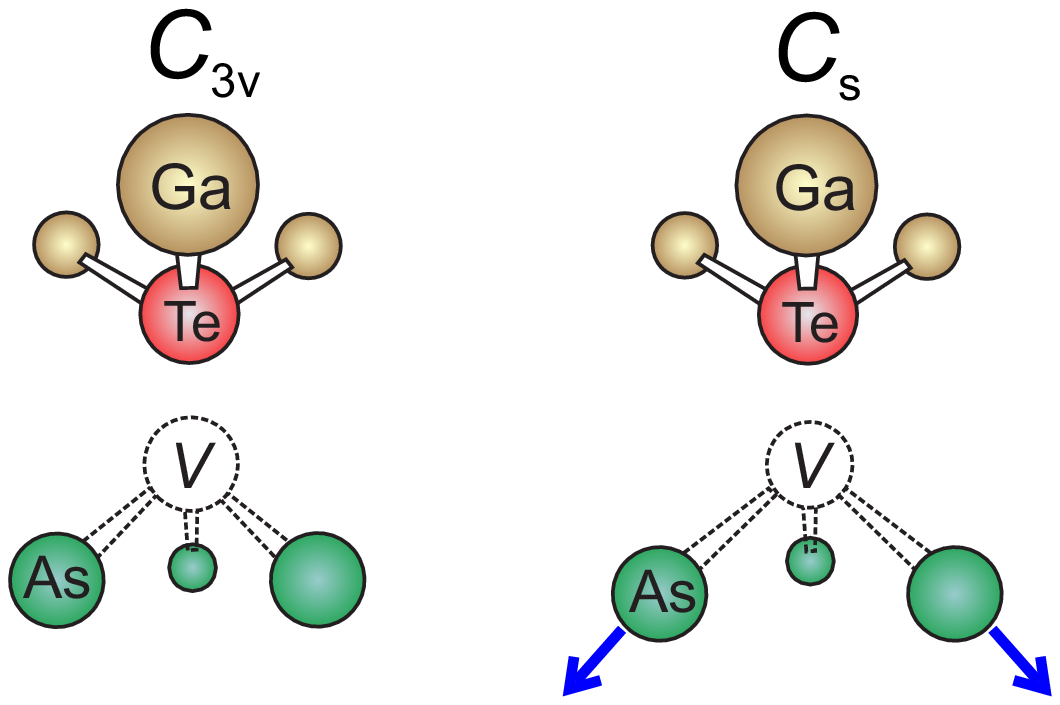}
\caption{\label{vgta}
Schematic of the GaAs:$V_\text{Ga}$-$\text{Te}_\text{As}$ defect, in the undistorted ($C_{3v}$ symmetry) and distorted ($C_s$ symmetry) structures.
In the distorted structure, two As atoms move away from the Ga vacancy (blue arrows) and the remaining one moves closer.}
\end{figure}

The involvement of donor species in the complex was first suggested by
Williams \cite{Williams1968} who noted a small shift of the emission
line depending on the type of the donor. For Te donors, luminescence
occurs at 1.18 eV. Subsequently, it was almost universally accepted
that the 1.18 eV line in Te-doped GaAs is due to the
$V_\text{Ga}$-$\text{Te}_\text{As}$ complex
\cite{Glinchuk1977,Glinchuk1976,Vorobkalo1973,Gutkin2000,Gutkin1994}. This
complex in Te-doped GaAs has been proposed to be responsible for the
compensation of Te$_\text{As}$ donors in GaAs
\cite{Wuyts1992,Gebauer2003}.
A schematic of the structure of the $V_\text{Ga}$-$\text{Te}_\text{As}$ complex is shown in Fig.~\ref{vgta}.

Based on a careful investigation of the 1.18 eV line in GaAs for
different $n$-type doping conditions, Glinchuk {\it et al.}
\cite{Glinchuk1977} proposed that the luminescence line corresponds to
the $(0/-)$ charge-state transition.
For donor concentrations $n > 10^{17}$cm$^{-3}$ optical emission was
caused by capture of free electrons, as evidenced from the almost
linear increase of capture rates $r=C_nn$ and corresponding decrease
of optical lifetimes according to Eq.~(\ref{tau}). An electron capture
coefficient $C_n=6.5\times10^{-13}$ cm$^3$s$^{-1}$ was deduced in that
work \cite{Glinchuk1977}.

Although the attribution of the 1.18 eV line to the
$V_\text{Ga}$-$\text{Te}_\text{As}$ defect seemed firm, the above
conclusions about the mechanism of luminescence were challenged by the
theoretical work of Baraff and Schl\"{u}ter \cite{Baraff1985}: using
the Green's function implementation of DFT in the local density
approximation (LDA), they found that an isolated $V_{\text{Ga}}$ can
occur in four charge states: 0, $-1$, $-2$, and $-3$. This suggests
that the $V_\text{Ga}$-$\text{Te}_\text{As}$ complex may have more
charge states than just 0 and $-1$ assumed by Glinchuk {\it et al.}
\cite{Glinchuk1977}. This conclusion is confirmed by our
first-principles calculations based on the HSE hybrid functional:
$V_\text{Ga}$-$\text{Te}_\text{As}$ indeed has three stable charge
states (0, $-$, and 2$-$, and thus two charge-state transition levels
in the band gap: $(0/-)$ at 1.23 below the CBM, and $(-/2-)$ at 1.03
eV below the CBM. Our calculated optical transition energies $
E_{\rm opt}$ [based on a configuration coordinate diagram as
illustrated in Fig.~1(b) of the main text] are 1.01 eV for the $(0/-)$
transition and 0.89 for $(-/2-)$. The result for the $(0/-)$
transition is within 0.2 eV of the experimental value of 1.18 eV,
within the error bar that has been empirically observed for other
defects when calculated with hybrid density functional theory
\cite{Lyons2010,Alkauskas2012}. Thus, within the limitations of the
theoretical approach, calculations support the assignment of the 1.18
eV band to a $A^0+e^-\rightarrow A^-$ transition.

In $n$-type material the stable charge state of the defect is 2$-$.
For a transition $A^0+e^-\rightarrow A^-$ to occur, the defect must
capture two holes first.  This should be reflected in the defect
emission being quadratically dependent on photo-excitation intensity
for small intensities \cite{Reshchikov2005}. However, in the case of
the 1.18 eV band this has not been observed.
The reason for this is likely that nonradiative capture of holes
suppresses the $A^-+e^-\rightarrow A^{-2}$ transition at even smaller
photo-excited carrier densities than used in experiments of
Refs.~\onlinecite{Glinchuk1976,Glinchuk1977,Gutkin1994,Gutkin2000}.
This is because radiative capture of an electron into the $-$ charge
state is reduced due to coulomb repulsion, and thus an
extremely small hole concentration (and therefore extremely low
excitation power) is required so that the defect in the $-$ charge
state does not rapidly transition to the neutral charge state by
nonradiatively capturing a hole.  When this ``saturation'' occurs, the
intensity of the $A^0+e^-\rightarrow A^-$ transition depends linearly
on the photo-excited carrier density.
Therefore, we conclude that $A^0+e^-\rightarrow A^-$ is responsible
for the observed 1.18 eV line.

\subsection{GaN:C$_\text{N}$}

Carbon is a common unintentional impurity in group-III nitride
materials. In some cases it is also introduced intentionally to
produce semi-insulating material. It has been shown to be one of the
sources of the notorious yellow luminescence (YL) in GaN
\cite{Ogino1980,Seager2004,Lyons2010}. Recent calculations
\cite{Lyons2010,Lyons2014} have demonstrated that GaN:C$_\text{N}$ is
a deep acceptor with emission energy around 2.14 eV, in good agreement
with the 2.2 eV peak observed in experiment
\cite{Ogino1980,Seager2004}. YL occurs when the electron in the
conduction band (or at a shallow donor) is captured by a neutral
carbon, C$_{\text{N}}^0$. We have previously reported first-principles
calculations of the luminescence lineshape pertaining to this
transition \cite{Alkauskas2012}; the calculated lineshape, as well as
effective vibrational parameters that characterize this lineshape, are
in an excellent agreement with experiment. In addition, the
nonradiative hole capture by a negatively charged carbon,
C$_{\text{N}}^{-}$, has been calculated \cite{Alkauskas2014}. In
$n$-type samples this process must precede the radiative capture of
electrons. Our calculated nonradiative capture coefficient $C_p$ was
in very good agreement with experimental data by Reshchikov
\cite{Reshchikov2014}. Thus, in the case of GaN:C$_{\text{N}}$, theory
and experiment agree with each other regarding: (i) the energy of the
radiative transition; (ii) the luminescence lineshape; (iii)
nonradiative capture coefficient $C_p$. All of these factors reinforce
the attribution of the observed YL in C-containing samples to
C$_{\text{N}}$.

\section{Electronic and atomic structure of the defects}

\subsection{GaAs:$V_\text{Ga}$-$\text{Te}_\text{As}$}

We start by considering the neutral charge state of a
$V_\text{Ga}$-$\text{Te}_\text{As}$ complex.  First, we fix all atoms
in unperturbed bulk positions.  The defect then has a $C_{3v}$
point-group symmetry, where the threefold symmetry axis goes through
the site of the vacancy and the neighboring $\text{Te}_\text{As}$ atom
(Fig.~\ref{vgta}). In this undistorted configuration there are two
degenerate single-particle states of $e$ symmetry in the band gap.  In
the neutral charge state these are filled with two electrons. As per
Hund's rule the triplet state $^3 A_2$ is the ground state of the
defect. Let us label multi-electron states with the occupation of
single-particle states $|e_{x}\bar{e}_{x}e_{y}\bar{e}_{y}\rangle$,
where ``bar'' denotes spin-down electrons. The triplet has three
components: $|1010\rangle$, $(|1001\rangle+|0110\rangle)/\sqrt{2}$,
and $|0101\rangle$, corresponding to spin projections $m_s=+1,$ $0$,
and $-1$.

When we allow the geometry to relax, the symmetry is lowered to
$C_s$. The distortion decreases one $V_\text{Ga}$-$\text{As}$ distance
and increases two others (Fig.~\ref{vgta}). The symmetry plane of the
defect passes through the site of the vacancy, the
$\text{Te}_\text{As}$ atom, and the As atom that has moved inward. The
single-particle $e$ states split into $a'$ and $a''$ components. In
the $C_s$ point group multi-electron states are given by specifying
the occupations of single-particle states
$|a'\bar{a'}a''\bar{a''}\rangle$. Like before, the triplet has three
components, which can be explicitly written as $|1010\rangle$,
$(|1001\rangle+|0110\rangle)/\sqrt{2}$, and $|0101\rangle$.  The
ground state of the defect in its neutral state is $^3A''$. Note that
the defect possesses three local minima, since the distortion can
occur in three different directions. This is not a Jahn-Teller
distortion, as assumed in Refs.~\cite{Gutkin1997,Gutkin2000}, since
the electronic ground state is not degenerate ($^3A_2$) in the
undistorted symmetric configuration. Rather, the lowering of the
symmetry of the defect should be considered polaronic, since the
driving force is not the breaking of an electronic degeneracy, but
rather the lowering of the electronic energy due to partial
localization of electronic wavefunctions.

Next we consider the negative charge state. If we consider the
undistorted configuration of the defect again, the ground state is an
orbital doublet $^2E$ with four electronic states: $|0111\rangle$,
$|1011\rangle$, $|1101\rangle$, and $|1110\rangle$. Because of the
electronic degeneracy, the system is unstable with respect to a
Jahn-Teller distortion, and the symmetry is lowered to $C_s$. The
electronic level splits into $^2A'$ and $^2A''$. We find that the
$^2A''$ is the ground state with components
$|a'(1)\bar{a}'(1)a''(1)\bar{a}''(0)\rangle$ and
$|a'(1)\bar{a}'(1)a''(0)\bar{a}''(1)\rangle$. As in the case of the
neutral defect, we find that in the distorted configuration one of As
atoms moves towards the vacancy, while the other two move
outward. Thus, the negatively charged defect possesses three local
minima, each of which has exactly the same symmetry plane as the
corresponding neutral defect; starting from the ground-state geometry
of the neutral defect and adding one electron to the system, we end up
with the defect in the negative charge state having the same symmetry
plane.

From the above analysis it follows that the ground state of the defect
in the neutral charge state is $^3A''$, while that in the negatively
charge state is $^2A''$. In an optical transition an electron in the
conduction band is captured into the defect state $a'$, which is
symmetric with respect to the reflection in the symmetry plane. The
ground-state geometries of the two charge states are different, but
both have $C_s$ symmetry with the same symmetry axis, and thus the
distortion includes only fully-symmetric $a'$ phonons. The latter
result is a prerequisite to the use of the Franck-Condon approximation
in optical transitions.

\subsection{GaN:C$_\text{N}$}

When a nitrogen atom is replaced by a carbon atom in the undistorted
GaN lattice, defect levels appear in the band gap. In the zinc-blende
structure there would be three degenerate $t_2$ levels made of carbon
$2p$ orbitals (with an admixture of N $2p$ orbitals on the
next-nearest neighbors). In the wurtzite structure the symmetry is
lowered from $T_d$ (tetrahedral) to $C_{3v}$. The $t_2$ orbitals are
split into $a_1$ and two degenerate $e$ orbitals.

In the negative charge state the three orbitals are filled with 6
electrons and the electronic configuration is $a_1^2e^4$. When we allow the
geometry to relax, the negative charge state retains the $C_{3v}$
symmetry, and the total wavefunction is of $^1A_1$ symmetry.
In the neutral charge state, there are 5 electrons to fill the defect
states in the band
gap.
The ordering of these $a_1$ and $e$ levels is not known {\it a
  priori}, but has to be determined from calculations. Depending on
the ordering of the levels, the ground state of C$_{\text{N}}^0$ in
the undistorted $C_{3v}$ geometry is either $^2E$ (configuration
$a_1^2e^3$) or $^2A_1$ (configuration $e^4a_1$). These two
multi-electron states are close in energy, thus, strictly speaking,
C$_{\text{N}}^0$ should be classified as a Jahn-Teller (if $^2E$ is
lower in energy) or pseudo-Jahn-Teller system (if $^2A_1$ is lower in
energy). From the electronic structure point of view, the system is
close to the triply-degenerate $^2T_2$ Jahn-Teller system of the
tetrahedral symmetry.

When we allow the geometry to relax, we find that the system can relax
into four local energy minima. One of them corresponds to the $C_{3v}$
symmetry, whereby the axial C-N bond is significantly elongated (2.11
\AA), while the other three retain a bond-length very similar to that
of the bulk Ga-N bond length (1.96 \AA). The electronic state is of
$^2A_1$ symmetry. In the other three cases the symmetry is lowered to
$C_s$, whereby one of the azimuthal C-N bonds is elongated, while the
other three remain close to bulk bond lengths. The ground state in
each of those configurations is of $^2A''$ symmetry.

We find that $^2A_1$ and $^2A''$ minima are
nearly-degenerate. Unfortunately, we find that the exact ordering of
these states depends very sensitively on various computational
parameters and cannot be determined accurately without the use of very
large supercells. For modest-size supercells the distance between
defects in the $xy$ plane is different from the distance in the $z$
direction, which causes an additional splitting between $a_1$ and $e$
orbitals beyond the natural crystal field of the wurtzite
lattice. However, the calculations consistently indicate that the
energy barrier between the minima is at least 0.1 eV. We thus expect
this system to behave as a static Jahn-Teller system, so the
Franck-Condon approximation and Born-Oppenheimer approximation
 apply to this defect, and we may use Eq.~(4) of the main text to
determine the capture coefficient. We choose the $^2A_1$ structure for
determining the momentum matrix element, though $\vert p_{if}\vert$
for $^2A''$ differs by less than 20 \%.

\section{Comparison between Hybrid and semilocal functionals}

In this section we comment on the role of hybrid functionals within
density functional theory in providing quantitative accuracy for the
capture coefficient $C_n$. When using traditional functionals, such as
the generalized gradient approximation (GGA), the band gap is severely
underestimated, and therefore the charge-state transition levels (as
well as $E_\text{opt}$) are also underestimated. In fact, for GaAs
calculations with the Perdew, Burke, and Ernzerhof (PBE) GGA
functional \cite{PBE} result in a band gap of less than 0.2 eV, and
therefore determining $E_\text{opt}$ is not possible since the
transition levels do not fall within the gap. Also, because of more
severe self-interaction errors in PBE, the defect wavefunctions are
more delocalized than when calculated with the hybrid functional of
Heyd, Scuseria, and Ernzerhof (HSE) \cite{HSE06}, leading to an
overestimation of momentum matrix elements.  HSE has also been shown
to accurately capture the conduction-band dispersion compared to
many-body perturbation theory techniques \cite{Dreyer2013,Yan2014}.

For GaN:C$_\text{N}$, a
fortuitous cancellation of these two errors occurs. Thus, the
calculated $C_{n}^{\text{PBE}}=0.8\times10^{-13}$cm$^3$s$^{-1}$ is
reasonably close to the HSE value, even though
$E_\text{opt}^{\text{PBE}}=0.95$ eV is severely underestimated. An
important advantage of the hybrid functional approach is that it
provides reliable first-principles predictions of \textit{both}
optical transition levels and capture coefficients, which is a crucial
requirement for defect identification.

We have also performed a test with another hybrid functional, namely
PBE0 \cite{Perdew1996,Carlo1999}.  We tuned the mixing parameter to
0.20 to reproduce the experimental band gap of GaN.  The resulting
momentum matrix element is very similar to the value obtained with HSE
(8\% larger).

\section{Quantum Defect  model for calculating capture rates}

Here we will compare the expression that we obtain for the radiative
capture coefficient [Eq.~(4) of the main text] with that obtained from
the ``quantum defect'' (QD) model \cite{Bebb1971, Bebb1969} that has
been used previously, as reviewed in Chapter 5 of
Ref.~\onlinecite{Ridley}. In the QD model, the defect potential near
the core is treated as a square well, while the long-range part has
the form of a Coulomb potential. The defect wavefunction is
assumed to be derived from valence-band states if it is an acceptor,
and conduction-band states for donors. It is important to note that
this happens to be a good approximation for the case-study defects in
this work, but will likely be a poor approximation in many other
cases, for instance if the wavefunction of an acceptor defect has
conduction-band character.

Quantities such as the effective Rydberg energy ($R_0^*$)
and effective Bohr radius ($a_0^*$) used below are characterized by
the valence-band effective mass $m^*_v$; of course, the relevant
effective mass for the density of states from which the electron
originates is that of the conduction band $m^*_c$.

\begin{figure}[h]
\includegraphics[width=8cm]{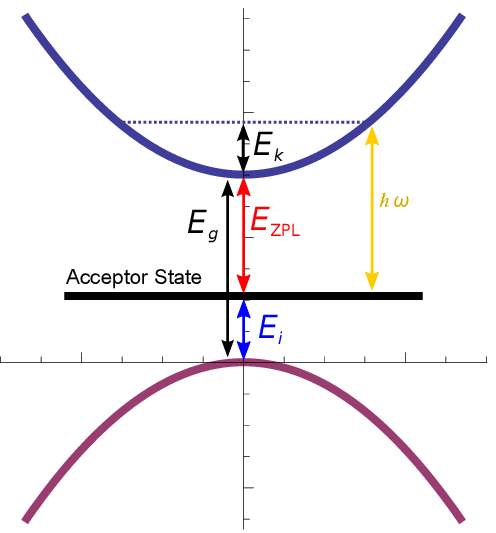}
\caption{\label{diag} Schematic of the energies involved in
  calculating the capture process based on the quantum-defect model.}
\end{figure}

As in the main text, we will consider capture of an electron from the
conduction band into an \emph{acceptor} state,
illustrated in Fig.~\ref{diag}. The conduction and valence bands are
assumed to be parabolic. The acceptor level has an energy $E_i$ above
the valence-band maximum (VBM) and $E_{\text{ZPL}}$=$E_g - E_i$ below
the conduction-band minimum (CBM), where $E_g$ is the band
gap. (cf. Fig.~1 in the main text).  In the QD model, $E_i$ is
referred to as the ionization energy.
The photoionization cross section for the process $A^{-} \rightarrow A^{0} + e^{-} $ is given by [Eq.~(5.91) of Ref.~\onlinecite{Ridley}]
\begin{equation}
\label{QD}
\sigma^{\text{QD}}(\hbar\omega)=\frac{16\times2^{2\mu}}{3}\alpha a_0^2\frac{R_0}{\hbar\omega}\frac{\vert p_{cv}\vert^2}{2m}\frac{1}{n_r}\frac{2\pi(\nu_{\text{T}}a^*_0)^3\Gamma^2(\mu+1)}{\Gamma(2\mu+1)}\frac{\left(\sin[(\mu+1)\tan^{-1}(E_{\textbf{k}}/E_i)^{1/2}]\right)^2}{(E_{\textbf{k}}/E_i)[1+(E_{\textbf{k}}/E_i)]^{\mu+1}}\left(\frac{2m^*_c}{\hbar^2}\right)^{3/2}E_{\textbf{k}}^{1/2},
\end{equation}
where $\alpha$ is the fine structure constant, $a_0$ is the Bohr
radius, $R_0$ is the Rydberg energy, $n_r$ is the refractive index,
$\vert p_{cv}\vert^2$ is the magnitude squared of the momentum matrix
element between the VBM and CBM,  $m$ is the free electron mass,
$\hbar\omega$ and $E_{\textbf{k}}$ are defined in Fig.~\ref{diag},
$\nu_{\text{T}}=\sqrt{R_0^*/E_i}$, and $\mu=Z\nu_{\text{T}}$, where
$Z$ is the charge of the defect.
[Note that in Eq.~(5.91) of Ref.~\onlinecite{Ridley}, the sine term is
not squared, which we attribute to a misprint in that work].
$p_{cv}$ should be distinguished from $p_{if}$ in the main text, which is the matrix element between the CBM and the defect
state; see below for discussion.

For a neutral defect ($Z=\mu=0$),
which is the relevant case for  capture of electrons at
the neutral deep acceptors we consider in this work, the
photoionization cross section is
\begin{align}
\label{sigQD1}
\sigma^{\text{QD}}_{Z=\mu=0}(\hbar\omega)
&=\frac{16}{3}\alpha a_{0}^2\frac{R_0}{\hbar\omega}\frac{1}{n_r}\frac{\vert p_{cv}\vert^2}{2m}\frac{2\pi (\nu_{\text{T}}a_0^*)^3}{(1+E_{\textbf{k}}/E_{\text{ZPL}})^2}\left(\frac{2m^*_c}{\hbar^2}\right)^{3/2}E_{\textbf{k}}^{1/2}
  \\
  \label{sigQD2}
\end{align}

We can convert
Eq.~(\ref{sigQD1}) into a capture coefficient by replacing the density
of electron states with photon states and the photon group velocity
with the electron band velocity (thus obtaining the capture cross
section), and finally multiplying by the band velocity to obtain
\cite{Ridley}:
\begin{align}
  C_{n}^{\text{QD}}(\hbar\omega)&=
\frac{16}{3}\alpha a_{0}^2\frac{R_0}{\hbar\omega}\frac{\eta_{\text{sp}}}{n_r}\frac{\vert p_{cv}\vert^2}{2m}\frac{2\pi (\nu_{\text{T}}a_0^*)^3}{(1+E_{\textbf{k}}/E_{\text{ZPL}})^2}\left(\frac{2m^*_c}{\hbar^2}\right)^{3/2}E_{\textbf{k}}^{1/2}\times\frac{(\hbar\omega)^2}{2m^*_cE_{\textbf{k}}(c/n_r)^2}\times\sqrt{\frac{2E_{\textbf{k}}}{m^*_c}}.
\end{align}
where we have also included the factor that accounts for the spin
selection rules $\eta_{\text{sp}}$ ($\eta_{\text{sp}}=0.5$ for our
test cases). We consider the capture coefficient for an electron originating from the CBM (corresponding to the threshold for photoionization), i.e., $\hbar\omega=E_{\text{ZPL}}$ and $E_{\textbf{k}}=\hbar\omega-E_{\text{ZPL}}=0$:
\begin{equation}
  \begin{split}
\label{QDfin}
  C_{n}^{\text{QD}}&=8n_r\eta_{\text{sp}}\frac{c}{3a_0(mc^2)^2}\frac{\vert p_{cv}\vert^2}{2m}E_{\text{ZPL}}\left[\pi(\nu_{\text{T}}a_0^*)^3\right]
  \\
  &=\frac{\eta_{\text{sp}}e^2n_r}{3m^2\varepsilon_0\pi c^3\hbar^2}\vert p_{cv}\vert^2 E_{\text{ZPL}}\left[4\pi(\nu_{\text{T}}a_0^*)^3\right] \, ,
  \end{split}
\end{equation}
Comparing this expression to Eq.~(4) of the main text
highlights two key approximations in the QD model. First, the model
does not account for the coupling of the defect with the lattice,
since the energy of the transition is taken to be $E_{\text{ZPL}}$
(instead of $E_{\text{opt}}$) which does not include the Frank-Condon
relaxation energy.  The second key approximation of the QD model is
that the momentum matrix element between the conduction band and
defect state is taken to be the one between the conduction and valence
bands; in place of $\widetilde{V}$ of Eq.~(4) of the main text (the
supercell volume in which $\textbf{p}_{if}$ is calculated),
$\vert p_{cv}\vert^2$ is multiplied by an effective volume of the
defect $V_{\text{eff}}=4\pi(\nu_{\text{T}}a^*_0)^3$.

We quantitatively compare Eq.~(\ref{QDfin}) to our first-principles
results using the parameters for our defects and host materials listed
in Table \ref{par}. We see that the QD model predicts a capture
coefficient for GaAs:$V_\text{Ga}$-$\text{Te}_\text{As}$ that is overestimated compared to
experiment, whereas the capture coefficient for GaN:C$_\text{N}$ is smaller than the experimental
range.

\begin{table}[h]
\caption{
Parameters for defects and host materials used in Eq.~(\ref{QDfin}), and the resulting  effective defect volume and electronic capture coefficient.
}
\begin{ruledtabular}
\label{par}
\begin{tabular}{c|cccccccccc}
  &   $E_{\text{ZPL}}$ (eV)\footnotemark[1] &$m^*_{c}$ (m$_{\text{e}}$)\footnotemark[2] & $m^*_{\text{h}}$ ($m_{\text{e}}$)\footnotemark[2] & $\varepsilon_r$\footnotemark[2] &$n_r$&$\vert p_{cv}\vert^2/2m$ (eV)&$R^*_0$ (eV)&$a_0^*$(\AA)& $V_{\text{eff}}$ (\AA$^3$)&C$_n^{\text{QD}}$ (cm$^3/$s)\\  \hline	
  GaAs:$V_\text{Ga}$-$\text{Te}_\text{As}$ & 1.23 &0.067 &0.5 &12.9&3.4\footnotemark[3] &6\footnotemark[1]  &0.04 &13.7&381&$10.15\times10^{-13}$\\
   GaN:C$_\text{N}$ & 2.48 & 0.20&1.0 &10.4&2.4\footnotemark[4]&4\footnotemark[5] &0.13 &5.5&66&$0.39\times10^{-13}$\\
\end{tabular}
\end{ruledtabular}
\footnotetext[1]{This work.}
\footnotetext[2]{Ref.~\onlinecite{Madelung}}
\footnotetext[3]{Ref.~\onlinecite{Skauli2003}}
\footnotetext[4]{Ref.~\onlinecite{Takahiro1997}}
\footnotetext[5]{Ref.~\onlinecite{Dreyer2013}}
\end{table}

As mentioned previously, the effective defect volume in the QD model
is given by
$V_{\text{eff}}=4\pi(\nu_{\text{T}}a^*_0)^3$. Ridley
\cite{Ridley} proposed a form of the photoionization cross section
that instead takes this volume to be an adjustable parameter. While
useful for, e.g., fitting photoionization or photoluminescence
spectra, this further reduces the predictive power of the model.

\bibliography{RadRecomb}